\documentclass[aps,prl,twocolumn,showpacs,superscriptaddress,groupedaddress]{revtex4-1}  
\usepackage{graphicx}  
\usepackage{dcolumn}   
\usepackage{bm}        
\usepackage{amssymb}   
\usepackage{amsmath}   

\begin{document}
\widetext
\begin{flushright}
TKYNT-11-10, RIKEN-QHP-8, BI-TP-2011/23
\end{flushright}

\title{Complex Heavy-Quark Potential at Finite Temperature from Lattice QCD }

\author{Alexander~Rothkopf$^{1,2}$, Tetsuo~Hatsuda$^{1,3}$, Shoichi~Sasaki$^1$}
\affiliation{$^1$Department of Physics, The University of Tokyo, Tokyo 113-0031, Japan}
\affiliation{$^2$Fakult\"{a}t f\"{u}r Physik, Universit\"{a}t Bielefeld, D-33615 Bielefeld, Germany}
\affiliation{$^3$Theoretical Research Division, Nishina Center, RIKEN, Saitama 351-0198, Japan}

\date{\today}

\begin{abstract}
We calculate for the first time the complex potential between a heavy quark and antiquark at finite temperature across the deconfinement transition in lattice QCD. The real and imaginary part of the potential at each separation distance $r$ is obtained from the spectral function of the thermal Wilson loop. We confirm the existence of an imaginary part above the critical temperature $T_C$, which grows as a function of $r$ and underscores the importance of collisions with the gluonic environment for the  melting of heavy quarkonia in the quark-gluon-plasma.
\end{abstract}

\maketitle

Heavy-quark bound states ($Q\bar{Q}$) are essential tools in the experimental and theoretical investigation of the high temperature state of QCD matter, the quark-gluon plasma  (QGP) \cite{Sarkar:2010zza}. In particular, 
  the suppression of heavy quarkonia such as $J/\psi$
   (the ground state of $c\bar{c}$ in the vector channel)
 in relativistic heavy-ion collision experiments at the Relativistic Heavy Ion Collider 
   \cite{Adare:2011yf} and at the Large Hadron Collider \cite{Aad:2010px}, provides us with an intriguing signal of the QGP.
  
In order to extract the physics of the QGP from
 heavy quarkonia, an intuitive as well as quantitative 
understanding of the involved physics is necessary: 
 In the classic approach of using a 
   Schr\"{o}dinger equation with a phenomenological $Q\bar{Q}$
    potential, Debye screening of color charges 
   is responsible for the melting of $Q\bar{Q}$  bound states
  above the deconfinement temperature $T_C$ \cite{Matsui:1986dk}. 
  A perturbative evaluation of the thermal Wilson loop
    using the hard-thermal-loop resummation technique
   shows that the static potential at high
      temperature  has an imaginary part induced by Landau damping 
      \cite{Laine:2006ns,Beraudo:2007ky}.
  An approach based on heavy-quark 
   effective field theory at finite temperature indicates
  an additional contribution to the imaginary part of the static 
  potential \cite{Brambilla:2008cx}. 
 In addition, lattice QCD simulations of the spectral function of charmonia
    at finite temperature show that   
    $J/\psi$ may survive even up to $ 1.5 T_C $ \cite{Asakawa:2003re}.  
    However, the connection among the different approaches 
    is not yet clearly understood.
      
The main purpose of this Letter is to unify the approach based on
the nonrelativistic Schr\"{o}dinger equation and that based on a
spectral decomposition of the $Q\bar{Q}$ correlator.
Utilizing lattice QCD simulations 
 of the medium surrounding the $Q\bar{Q}$, we extract a nonperturbative  potential at 
 \textit{any temperature}, especially in the phenomenologically important and hitherto inaccessible region around $T_C$. 

\begin{figure*}[th!]
\includegraphics[scale=0.3,angle=-90]{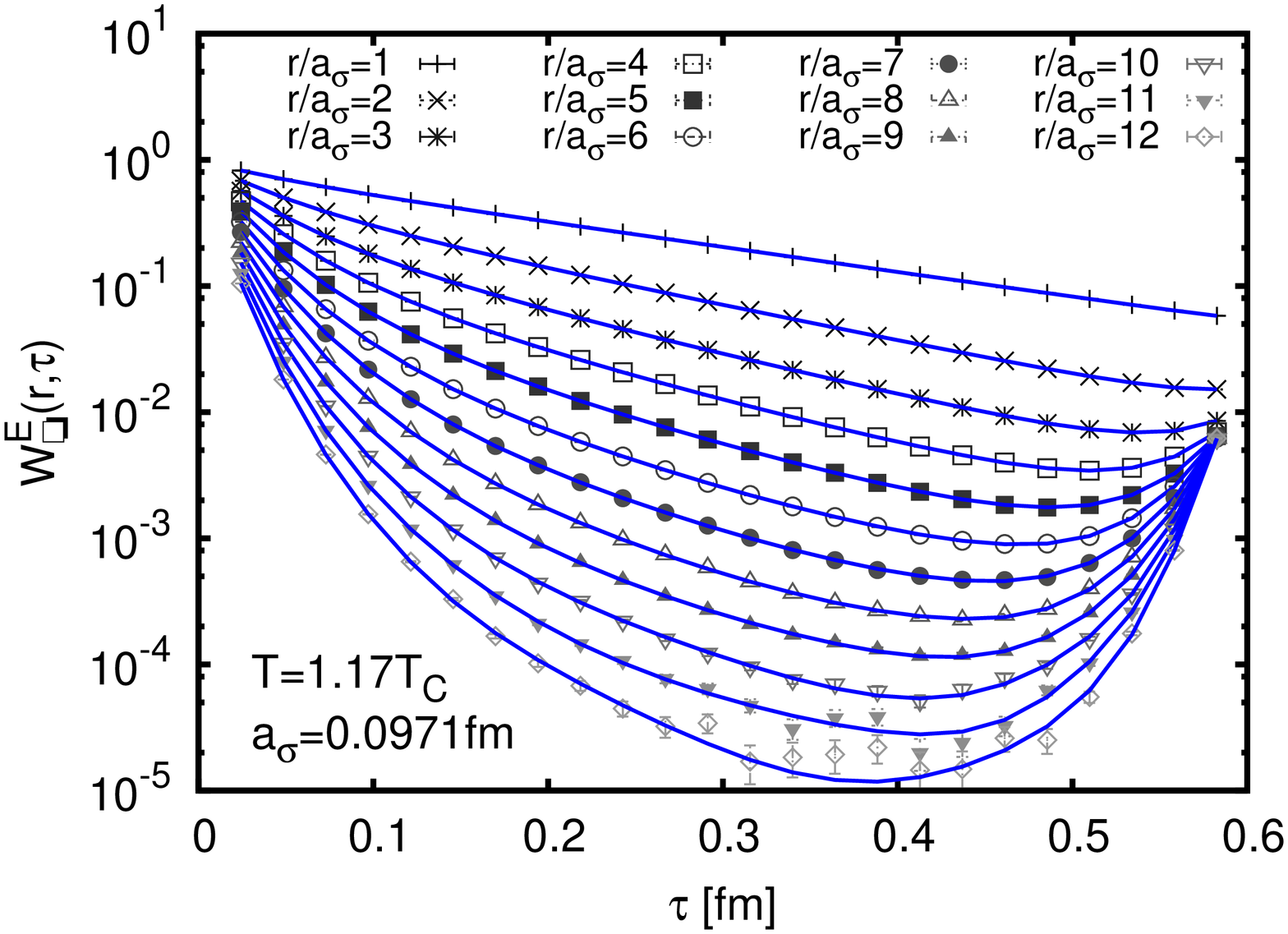}\hspace{0.8cm}
\includegraphics[scale=0.3,angle=-90]{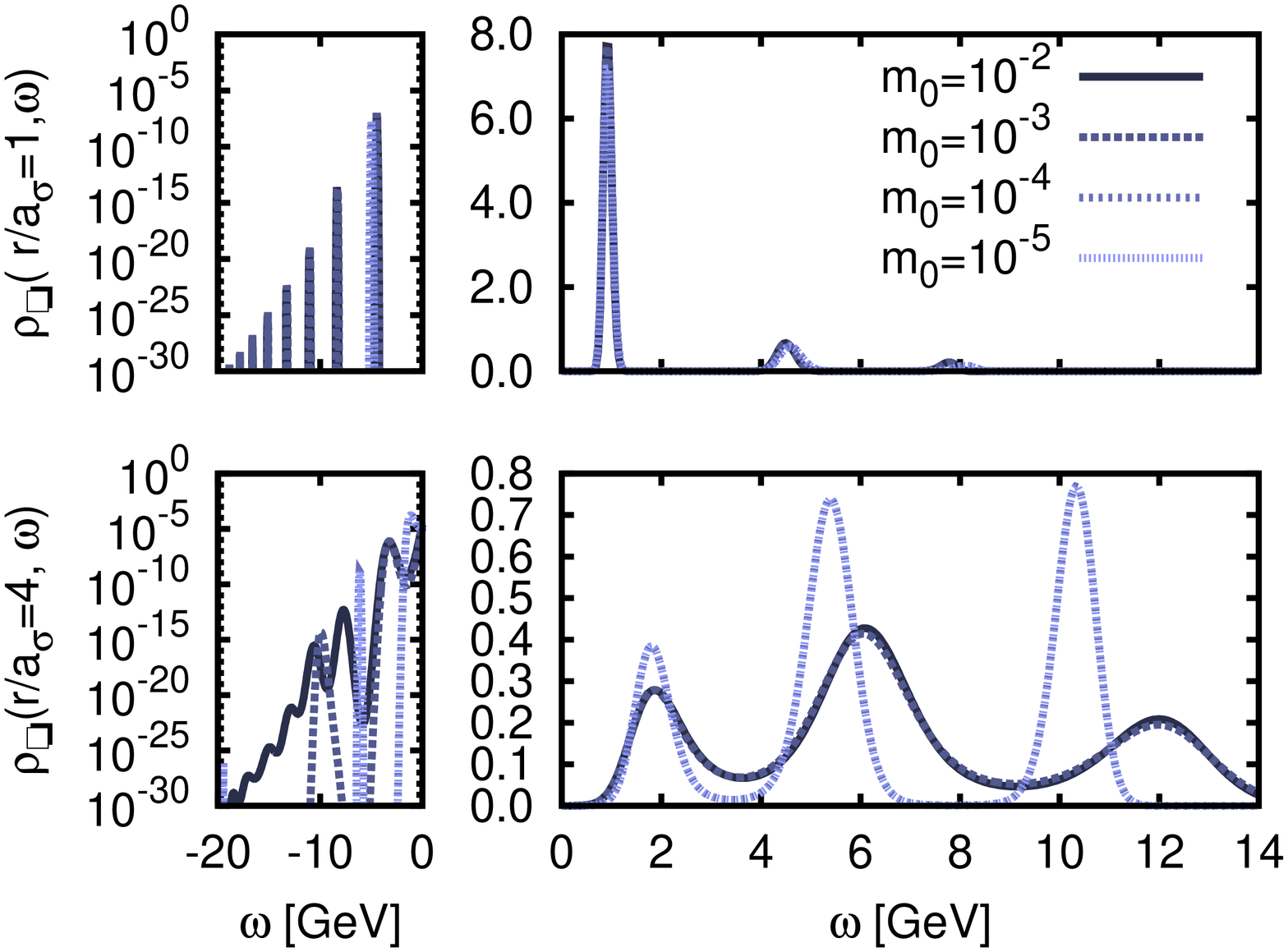}
\caption{(Left) Lattice data for the Euclidean thermal  Wilson-loop as a function of the 
 imaginary time $\tau$ at $T/T_C=1.17$. The solid lines show the 
 results obtained after the MEM reconstruction of the spectral function.
 (Right) Spectral functions and their prior dependence
  obtained by the MEM with the thermal Wilson-loop data 
 at   $r/a_{\sigma}=1, 4$.   Positive $\omega$ regions and negative $\omega$ regions are
   plotted separately with a different vertical scale.}
\label{Fig:WL-Spec}
\vspace{-0.35cm}
\end{figure*}

The starting point for deriving the effective Schr\"{o}dinger equation at finite $T$
is the real-time forward $Q\bar{Q}$ correlator ($t > 0$) \cite{Laine:2006ns}
\begin{eqnarray}
D^>({\bf r},t)
=\Big\langle M({\bf x},{\bf y},t) M^\dagger({\bf x},{\bf y},0) \Big\rangle , 
\label{Eq:ForwProp}
\end{eqnarray}
where
$M({\bf x},{\bf y},t)=\bar{Q}({\bf x},t)\gamma U[{\bf x},{\bf y}]Q({\bf y},t)$
with $\gamma$ denoting gamma matrices ($\gamma=\gamma^\mu$ for $J/\psi$)
 and $U[{\bf x},{\bf y}]$ being a straight Wilson line connecting 
  $({\bf x},t)$ and $({\bf y},t)$.  The relative coordinate is defined as
  ${\bf r}={\bf x}-{\bf y}$.
 The angular brackets impliy the thermal average.
  In the nonrelativistic limit, where the heavy-quark mass $m_Q$ is much larger
 than any other scale of the system,
 one may rewrite Eq.(\ref{Eq:ForwProp})
  in terms of a path integral over the positions and momenta of the heavy quarks:
\begin{eqnarray}
 D^>_{\rm NR}({\bf r},t)
  & \propto & \int {\cal D}[{\bf z}_1,{\bf z}_2,{\bf p}_1,{\bf p}_2]
 \times \label{Eq:PathIntProp} \\
\nonumber & & \! \! \! \! {\rm exp}\Big[ i\int_{0}^{t} ds \sum_{j=1,2}\Big(  {\bf p}_j(s) 
\dot{{\bf z}}_j(s) - \frac{{\bf p}_j^2(s)}{2m_Q}\Big)\Big]
 \times W_{\Gamma} 
 \end{eqnarray}
with the Wilson loop,
$W_{\Gamma} = \left\langle {\rm exp}\Big[ - {ig}\int_{\Gamma} dx_\mu A^\mu(x) \Big] \right\rangle $. 
  Here $\Gamma$ denotes a closed loop in Minkowski space-time,
 defined by the path of the quark and antiquark.
    This formula  is a finite temperature generalization
 of the result in Ref. \cite{Barchielli:1986zs} and is valid up to $O(1/m_Q)$.

From here on, we focus on the leading part of the potential, obtained in the heavy-quark
 limit $m_Q\to\infty$.  In this case, it is enough to consider a rectangular 
 path for the Wilson loop $ W_\Gamma \rightarrow  W_\square(r=|{\bf z}_1-{\bf z}_2|,t)$
  with a Fourier decomposition;
\begin{equation}
 W_\square(r,t) = \int_{-\infty}^{+\infty}
  d\omega e^{-i\omega t} \rho_\square(r,\omega). \label{Eq:SpecFunc1}
\end{equation}
Equation (\ref{Eq:ForwProp}) for $m_Q \rightarrow \infty$ also has a spectral decomposition
 composed of three parts, $\rho_{\rm loop}$, $\rho_{\rm staple}$ and $\rho_{\rm handle}$:  
 one can show that  $\rho_\square(r,\omega)$ in 
  Eq.(\ref{Eq:SpecFunc1}) corresponds to $\rho_{\rm loop}$ 
  and is  positive semidefinite for all $r$ and $\omega$ \cite{Rothkopf:2009pk}.
 
Through Eq.(\ref{Eq:SpecFunc1}) we are led to the following 
nonperturbative definition of the in-medium potential 
\begin{equation}
 \frac{i\partial_t W_\square(r,t)}{W_\square(r,t)}
 =\frac{\int d\omega \, \omega\, e^{-i\omega t} \rho_\square(r,\omega)}
 {\int d\omega\, e^{-i\omega t} \rho_\square(r,\omega)} \equiv V_{\square}(r,t) .
\label{Eq:PropPot}
\end{equation}
Indeed, the solution of Eq.(\ref{Eq:PropPot}) yields the transfer matrix
 compatible relation $W_\square(r,t)= \exp \left[ -i \int_0^t V_{\square}(r,s)ds \right]$ 
 and  thus leads to an in-medium Schr\"{o}dinger equation
\begin{equation}
 i\partial_t D_{\rm NR}^>({\bf r},t)=
  \Big(-\frac{\nabla_r^2}{m_Q} +V_{\square}(r,t) \Big) D_{\rm NR}^>({\bf r},t) .
  \label{Eg:WLSchroed}
\end{equation}

\begin{figure*}[th!]
\includegraphics[scale=0.3,angle=-90]{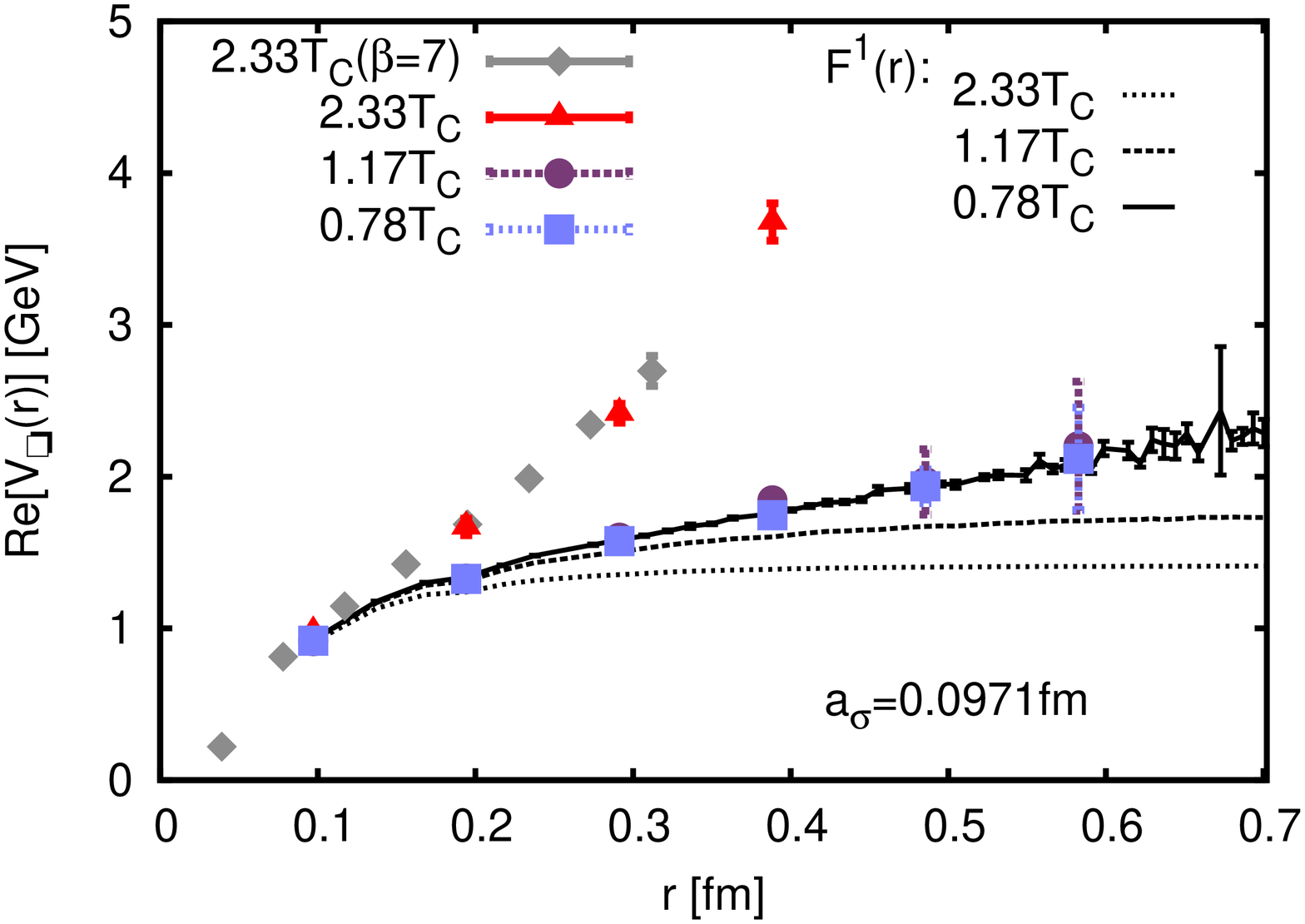}\hspace{0.8cm}
\includegraphics[scale=0.3,angle=-90]{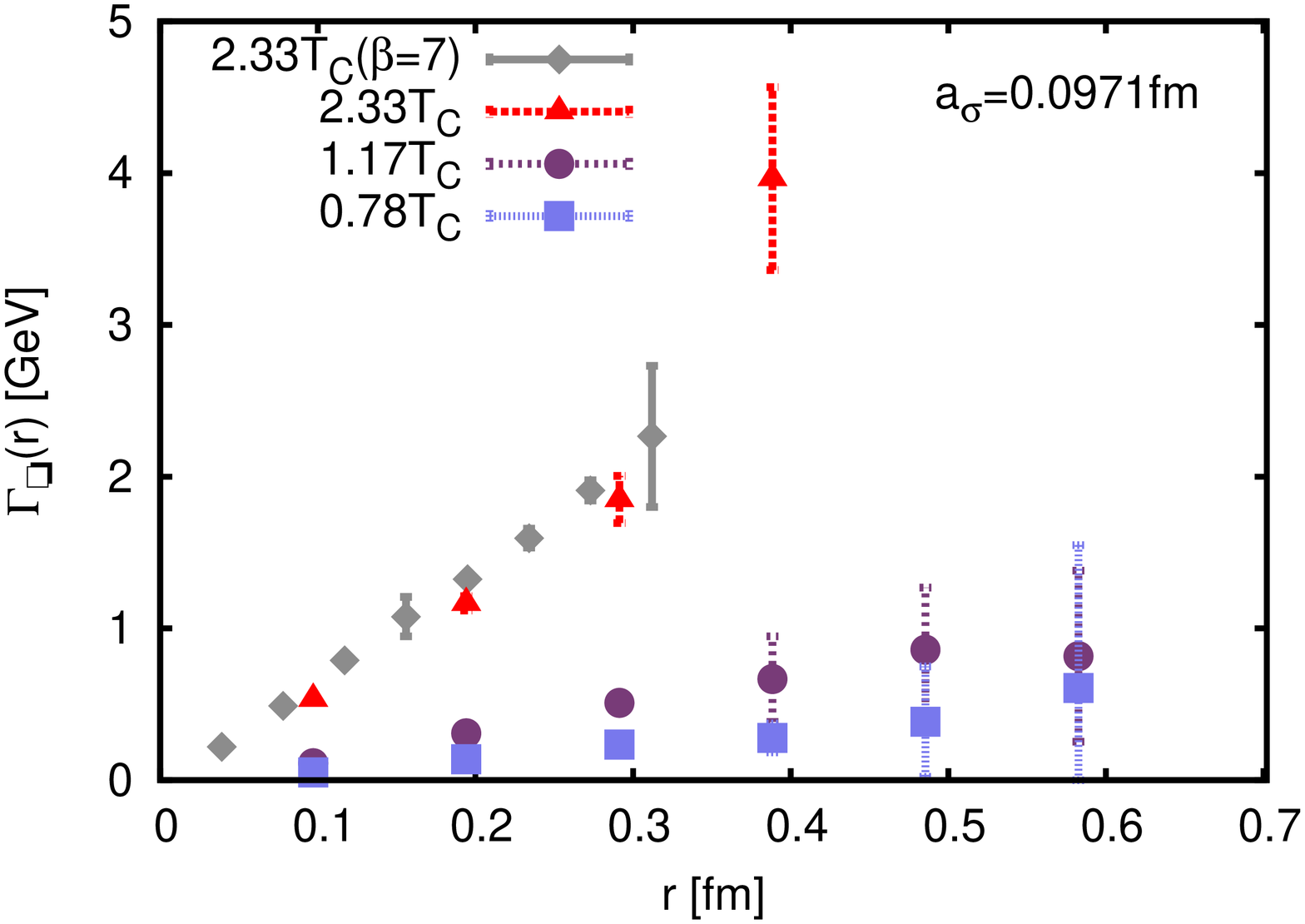}
\caption{(Left) The real part of the in-medium heavy-quark potential for three different
 temperatures across $T_C$ with $a_{\sigma}=0.0971$ fm (solid squares, circles and triangles).
 Solid and dashed lines are
the color-singlet free energies $F^1(r)$ 
 in the Coulomb gauge at corresponding temperatures.
  The result for the finer lattice ($a_{\sigma}=0.039$ fm at $T/T_C=2.33$ is also shown
  for comparison (solid diamonds).
  (Right)  The imaginary part of the potential, obtained from the width of the lowest lying peak
  of the spectral function. The same symbols are used as the left figure.}
\label{Fig:PotReIm}
\vspace{-0.4cm}
\end{figure*}

 In the following we  study the 
 peaks of the spectral function in Eq.(\ref{Eq:PropPot}) 
 around $\omega =0 $, since it is the physics at late times $t$ that determines the potential.
 Based on Eq.\eqref{Eq:PropPot}, the real part of the potential is thus given by the peak position, 
 whereas the imaginary part is related to the peak width:
 For example, if $\rho_\square(r,\omega)$ exhibits a Breit-Wigner peak with
 position $\omega_0(r)$ and width $\Gamma_0(r)$, 
 we obtain $V_{\square}(r,t)\rightarrow V^{\rm (BW)}(r) = \omega_0(r) + i\Gamma_0(r)$,
  while if it contains a single Gaussian peak, we have
$V_{\square}(r,t)\rightarrow V^{\rm (G)}(r,t) = \omega_0(r) + i\Gamma_0^2(r)\, t$.

 A nonperturbative method to determine $\rho_\square(r,\omega)$ and
 hence $V_{\square}(r,t)$ for all $T$ is based on lattice QCD simulations.
 We first make an analytic continuation of Eq.\eqref{Eq:SpecFunc1}
  to imaginary time and connect the Euclidean thermal Wilson-loop to the spectral function through 
  the Laplace transform,
$ W^{\rm E}_\square(r,\tau) 
= \int d\omega e^{-\omega \tau}  \rho_\square(r, \omega)$.
 This quantity  is purely
 real and not symmetric under the reflection $\tau \leftrightarrow \beta -\tau$
  by definition.  As shown in Ref. \cite{Rothkopf:2009pk}, $\rho_\square(r,\omega)$
   in a finite volume 
 allows for a spectral decomposition in terms of delta-function peaks:
 \begin{equation}
 \rho_\square(r, \omega)
 = \sum_{n,n'} P_n(T)|A_{n,n'}(r)|^2 \delta (\omega - E_{n'}^{Q\bar{Q}}(r)+ E_n).
\label{Eq:SpecDec}
\end{equation}
 Here  $E_n$ is the energy eigenvalue of a state $|n \rangle$ 
   without $Q\bar{Q}$, while  $E_{n'}^{Q\bar{Q}}(r)$ is the eigenvalue
   (relative to $2m_Q$) of a state $|n';r \rangle$ including a
   $Q\bar{Q}$ separated by a distance $r$. The matrix element and the
     Boltzmann factor are defined as
  $A_{nn'}(r) \equiv \sum_{{\bf y}}\langle n | M({\bf y}+{\bf r}, {\bf y}, t=0) |n' \rangle $ and
 $P_n(T) \equiv e^{ -E_n/T }/\left[ \sum_n e^{-E_n /T} \right]$, respectively.
  The peak positions of Eq.(\ref{Eq:SpecDec})
   do not depend on $T$ (since $E_n$ and $E_{n'}^{Q\bar{Q}}(r)$ are $T$-independent),
   while  the pole residues $P_n(T)|A_{n,n'}(r)|^2$
    depend on $T$ due to the Boltzmann factor.
 The various possible combinations of $n'$ and $n$ in 
 $\rho_\square(r, \omega)$ at finite $T$ lead to a number of peaks
 as large as  $N \times N$, if $\rho_\square(r, \omega)$ at $T=0$ has $N$ peaks.

  The position and width of a physical resonance in an {\em infinite} volume
   are obtained from the {\em envelope} of a bunch of 
   delta-function peaks in  Eq.(\ref{Eq:SpecDec}). 
  A suitable technique to extract such envelopes from
  lattice QCD data of $W^{\rm E}_\square(r,\tau)$ is
  the maximum entropy method (MEM)  \cite{Asakawa:2000tr}.
  In this Letter, we utilize a high precision MEM 
    with an extended search space developed by one of the present 
    authors \cite{Rothkopf:2011}.
   By separating the size of the search space from the number
   of data points, it allows a consistent comparison of spectra even if the 
   temporal extent of the underlying lattice is changed.

  The stability of the position and width of the resonance 
  reconstructed by the MEM
  can be checked through changing the amplitude of the prior and by increasing 
  the number of temporal data points \cite{Asakawa:2000tr}.
  In addition, the effect of a discrete lattice spectrum
  on the position and width of the resonance can be checked
   by controlling the number of peaks and their relative separation  
  under the variation of the lattice spacing $a$ and the lattice size $L$ 
  (see e.g. \cite{Yamazaki:2002} for the variation of $a$).
   
We perform quenched 
lattice QCD simulations of $W^{\rm E}_\square(r,\tau)$
 by using the simple plaquette gauge action 
on an anisotropic $20^3\times N_\tau$ lattice.
The anisotropy ratio
between the the spatial and temporal lattice spacing is taken to be
 $\xi \equiv a_{\sigma}/a_{\tau} = 4$. We fix $a_\sigma$ to be $0.097{\rm fm}$  ($\beta=6.1$)
  and adopt the temporal lattice sizes $N_\tau=36,24,12$  
 which correspond to temperatures $T/T_C=0.78, 1.17, 2.33$ with 
$T_C \simeq 290{\rm MeV}$\cite{Matsufuru:2001cp}. Our spatial lattice size $L\sim 2$fm can accommodate the characteristic $J/\psi$ scale of $r_{J/ \psi}\sim 0.5$fm. 
 After collecting the lattice data,
 we carry out the MEM  over a frequency interval
$I_\omega = [\omega_{\rm min}, \omega_{\rm max}]$ of $N_\omega=1500$ points. 
Our choice corresponds to $I_\omega\simeq[-21 {\rm GeV},
 42{\rm GeV}]$ at $N_\tau=24$. We use a prior distribution 
of the form $m(\omega)=\frac{1}{\omega+\omega_{0}}$,
motivated by the canonical dimension of $\rho_{\square}(r,\omega)$.
The parameter $\omega_0 (>0)$ is fixed by setting the amplitude
$m_0 = m(\omega_{\rm min})$ at the smallest frequency in $I_{\omega}$.

In Fig.\ref{Fig:WL-Spec}(left) we plot typical Wilson-loop data 
 as a function of $\tau$ for various distances $r$ at $T=1.17T_C$.
 Spectral functions obtained by the MEM at $r=a_{\sigma}\simeq 0.1$ fm
  and $r=4a_{\sigma}\simeq 0.39$ fm are plotted in Fig.\ref{Fig:WL-Spec}(right).  
  The falloff of $W_\square^{\rm E}(r,\tau)$ for small and intermediate values of 
   $\tau$ in the left figure 
   corresponds to the peaks located in the $\omega >0 $ region seen in the right figure:
  they arise from gluonic interactions between the temporal Wilson lines.
  The upward trend of $W_\square^{\rm E}(\tau,r)$ around $\tau=\beta$ on the other hand is induced by 
  extremely small structures located in the $\omega<0$ region:
  These arise from short distance gluon interactions connecting the spatial Wilson lines
  across the compactified temporal axis. 
  Although the spectral function in the negative $\omega$ region is important for reproducing
  $W_\square^{\rm E}(r,\tau\sim \beta)$ due to its  exponential ``enhancement" by the Laplace transform,  its effect on $V_\square(r,t)$  in 
  Eq.(\ref{Eq:PropPot}) is negligible due to its extremely small residue.


The real and imaginary parts of the potential are obtained from
fitting the lowest lying peak  with  a Breit-Wigner and Gaussian shape. The
identical results for position and width are shown in Fig.\ref{Fig:PotReIm}(left) 
and (right),
respectively. The error bars are obtained from the variance in the peak structure between different choices for
the amplitude of the prior distribution. The interval of $10^{-2}\ge m_0 \ge 10^{-6}$ is chosen to span as many 
orders without introducing numerical instabilities in the minimization process. We have checked that the standard 
MEM error, estimated from the stability of the spectral function given $m_0$ \cite{Asakawa:2000tr}, is much smaller than the error from the variation of $m_0$.

$\mathbf{T=0.78T_C}$: ${\rm Re}[V_{\square}(r)]$ denoted by the filled squares in 
 Fig.\ref{Fig:PotReIm}(left) 
is found to show a linearly rising potential at long distances, a result consistent
 with quark confinement below $T_C$. To study the short distance behavior
  of the potential, we have also carried out simulations with a $2.5$ times smaller
  lattice spacing 
  $(\beta=7, \xi=4, a_\sigma=0.039{\rm fm},20^3\times96)$ at the same $T/T_C$;
 the resulting  ${\rm Re}[V_{\square}(r)]$ can be fitted well by a Coulomb + linear
  potential \cite{footnote2}.
 Furthermore, our results agree with the color-singlet free energies $F^1(r)$ in the Coulomb
   gauge (solid black line in the left panel) at all measured distances $r$.
 Note that $\Gamma_{\square}(r)$ appears to be small and is consistent with zero
  within the statistical and systematic errors; see the filled squares in 
  Fig.\ref{Fig:PotReIm}(right). 

$\mathbf{T=1.12T_C}$: At this temperature  Re$[V_{\square}(r)]$ has apparently frozen at around the same strength found at $T=0.78T_C$, as shown by the filled circles in the left panel.
 This is in contrast to the behavior of $F^1(r)$ at the same $T$ (the long dashed curve
 in the left panel)
 which exhibits a significant thermal screening.
 As for $\Gamma_{\square}(r)$, there is a tendency to develop a nonzero value,
  which grows as $r$ increases; see the filled circles in the right panel. 

$\mathbf{T=2.33T_C}$: At the highest available temperature,
${\rm Re}[V_{\square}(r)]$ and $\Gamma_{\square}(r)$ exhibit a strong rise as a function of $r$,
shown by the filled triangles.
 To test the effects of a small number of temporal data points $N_{\tau}$ \cite{Asakawa:2000tr}
 and also whether the discrete lattice spectrum \cite{Yamazaki:2002}
   might blur the MEM image and thus might lead to an artificial
  broad peak, we compare the results of
 $(\beta=6.1,a_\sigma=0.097{\rm fm},20^3\times 12)$
and those of $(\beta=7,\xi=4,a_\sigma=0.039{\rm fm},20^3\times 32)$:
The latter has more temporal data points; hence, it has finer (coarser) 
resolution at low (high) frequency.
 The results shown by solid diamonds in Fig.\ref{Fig:PotReIm}(left) and (right) 
   are consistent with the solid triangles.
 This crosscheck indicates that the width broadening observed from the MEM
    is a physical effect \cite{footnote3}. We need, however,
  further  systematic studies to confirm this point by changing $N_{\tau}$, $a$ and 
  $L$ independently.

\begin{figure}[t]
\includegraphics[scale=0.3,angle=-90]{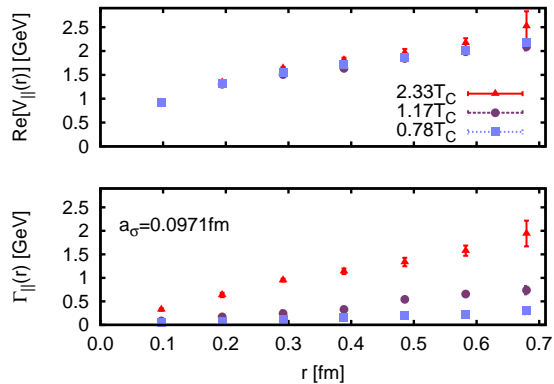}
\caption{The real part  (upper panel) and the imaginary part (lower panel) of the potential
 obtained from the Wilson lines $W^{\rm E}_{||}$.  }
\label{Fig:Spectrum}
\vspace{-0.4cm}
\end{figure}

 Although we construct our potential by starting from the 
 mesonic operator $M({\bf x},{\bf y},t)$ with a straight Wilson line
 in Eq.\eqref{Eq:ForwProp}, one has the freedom to choose 
 the operator $M$ differently and, hence, the forward correlator $D^>$ and the potential $V$.  
 On the other hand, observable quantities, such as 
the dilepton emission rate  must be independent of such differences \cite{Burnier:2007qm}. 
 Therefore, there must be a trade-off between the 
  real and the imaginary part of the potential to leave the 
   observables unchanged.  
  To study this point, we consider an operator
  $M$ with $U[{\bf x},{\bf y}]=1$ in the Coulomb-gauge. 
 In this case, we have ``Wilson lines" without the spatial link $W_{||}(r,t)$.
 In Fig.\ref{Fig:Spectrum}, we show the real and imaginary parts of the 
  potential ($V_{||}$) obtained from $W^{\rm E}_{||}(r,\tau)$.
 In the confinement phase below $T_C$,
  $V_{||}$ and $V_\square$ agree quite well.  On the other hand,
  in the deconfinement phase,
   both ${\rm Re}[V_{||}(r)]$ and $\Gamma_{||}(r)$
   exhibit a less pronounced rise in $r$.
   We observe that a weaker real part is accompanied by a weaker
   spectral width, which could be a sign of the trade-off mentioned above.
     This mechanism has to be made quantitative in future studies by solving the 
     time dependent Schr\"{o}dinger equation for an initial $Q\bar{Q}$ wave packet 
     entering the QGP.

We have presented a 
 nonperturbative derivation of the Schr\"{o}dinger equation for heavy quarkonia
  and a first evaluation of the corresponding {\em complex}
   in-medium potential, based on quenched lattice QCD. Our numerical results show that, even though the potential agrees with the color-singlet free energies below the phase transition, the correct physics above $T_C$ can be obtained only if the real and imaginary part are taken into account together. The temperature insensitive real part around $T_C$ suggests furthermore that the growth of the imaginary part, i.e. an increasing number of collisions with the medium, may
  play a more important role to destabilize $Q\bar{Q}$ than the screening effects  
  \cite{Morita:2007pt}. We have also discussed a possible mechanism to obtain the relevant physics independent of a particular choice of the underlying operator by balancing the real and imaginary part.
Our complex potential opens up new possibilities to study the dynamics of the QGP transition by providing first-principles input to nonrelativistic real-time simulations, going beyond both models and perturbation theory. 
   
Our ongoing work aims at full QCD simulations with dynamical fermions, since
 these additional degrees of freedom may affect both the real and imaginary part of the complex potential substantially.
In addition, larger and finer lattices are needed in order to assess the relative significance of Debye screening vs the collisional effects from short distance to long distance in more detail.

The authors thank T.Matsui, H.Satz and M. Laine for stimulating discussions, N. Brambilla and Y. Burnier for insight into pNRQCD, and Y. Maezawa for his support with the lattice setup. A.R. acknowledges support from MEXT and the Young Researchers Initiative of the supercomputing center at The University of Tokyo. A.R. and T.H. were 
 partially supported by Grant-in-Aid of MEXT No. 22340052.


\begin{thebibliography}{99}
\bibitem{footnote2}
Since $W_\square(r,\tau)$ at low $T$ is clearly dominated by a single exponential falloff, we used naive exponential fitting for $\beta=7$ in the determination of $V_\square(r)$ at small $r$.
\end{thebibliography}
\end{document}